\documentclass[aps,prl,twocolumn,showpacs,superscriptaddress,preprintnumbers,amsmath,amssymb]{revtex4-1}
\usepackage[breaklinks=true,colorlinks=true
,urlcolor=blue
,anchorcolor=blue
,citecolor=blue
,filecolor=blue
,linkcolor=blue
,menucolor=blue
,pagecolor=blue
,linktocpage=true
,pdfproducer=medialab
,pdfa=true
]{hyperref}
\usepackage{graphicx,hyperref,color}
\usepackage{color}
\usepackage[dvipsnames]{xcolor}
\usepackage{graphicx}
\usepackage{dcolumn}
\usepackage{bm}
\usepackage{bbm}
\usepackage{mathtools}

\usepackage{tikz}
\usetikzlibrary{arrows,arrows.meta,intersections, calc,positioning,decorations.pathreplacing,decorations.pathmorphing,shapes}
\usetikzlibrary{patterns}
\usetikzlibrary{decorations.markings}
\usetikzlibrary{knots}

\def\frac#1#2{{#1\over #2}}

\def\be{\begin{equation}}
\def\ee{\end{equation}}
\def\ba{\begin{eqnarray}}
\def\ea{\end{eqnarray}}
\def\de{\partial}

\def\la{\langle}
\def\lb{\rangle}

\usepackage{braket}

\begin{document}

\title{{\large \textbf{Essay: Emergent Holographic Spacetime from Quantum Information}}}
\preprint{YITP-25-58}

\author{Tadashi Takayanagi}
\affiliation{Center for Gravitational Physics and Quantum Information, Yukawa Institute for Theoretical Physics, Kyoto University, \\Kitashirakawa Oiwakecho, Sakyo-ku, Kyoto 606-8502, Japan}
\affiliation{Inamori Research Institute for Science, \\620 Suiginya-cho, Shimogyo-ku, Kyoto 600-8411, Japan}


\begin{abstract}
Holographic duality describes gravitational theories in terms of quantum many-body systems. In holography, quantum information theory provides a crucial tool that directly connects microscopic structures of these systems to the geometries of gravitational spacetimes. One manifestation is that the entanglement entropy in quantum many-body systems can be calculated from the area of an extremal surface in the corresponding gravitational spacetime. This implies that a gravitational spacetime can emerge from an enormous number of entangled qubits. In this Essay, I will discuss open problems in this area of research, considering recent developments and outlining future prospects towards a complete understanding of quantum gravity. The first step in this direction is to understand what kind of quantum circuits each holographic spacetime corresponds to, drawing on recent developments in quantum complexity theories and studying concrete examples of holography in string theory. Next, we should extend the concept of holography to general spacetimes, e.g., those spacetimes which appear in realistic cosmologies, by utilizing the connections between quantum information and holography. To address the fundamental question of how time emerges, I will propose the concepts of {\it pseudo-entropy} and {\it time-like entanglement} as a useful tool in our exploration.

\textit{Part of a series of Essays in Physical Review Letters which concisely present author visions for the future of their field.}

\end{abstract}

\maketitle



{\it From black hole to holography.}---One of the most important unsolved problems in physics is to fully construct and understand the microscopic theory of gravity, known as {\it quantum gravity}. This issue is crucial for answering a fundamental question about the natural world: How was the Universe created? Today, we know that three of the four fundamental forces in nature—the electromagnetic, weak, and strong forces— are described microscopically by gauge theories in quantum field theory. On the other hand, gravity, the remaining force, cannot be treated in the same way because gravitational interactions are not renormalizable. Interestingly, however, the special nature of gravity can be found even at the classical theory level. The most famous manifestation is that black holes possess gravitational entropy \cite{Bekenstein:1973ur,Hawking:1975vcx}. This means that the amount of information hidden inside a black hole can be measured by the black hole's entropy $S_{BH}$ given by the beautiful formula:
\be
S_{BH}=\frac{\mbox{A}(\Sigma)}{4G_N},  \label{BH}
\ee
in natural units, where $A(\Sigma)$ is the area of the black hole horizon $\Sigma$ and $G_N$ is the Newton constant. There are two surprising features. One is that black hole entropy is proportional to the area rather than volume, as generally expected from thermodynamics. The other one is that we can compute gravitational entropy even in the classical theory of general relativity. In familiar quantum field theories of a scalar, fermion, and gauge field, entropy is calculated only after quantizing the theory. Indeed, while solitons in ordinary quantum field theories cannot carry any entropy at the classical level, black hole solitons in general relativity can. This raises the question: Where does this entropy originate?

\begin{figure}[ttt]
   \centering
   \includegraphics[width=6cm]{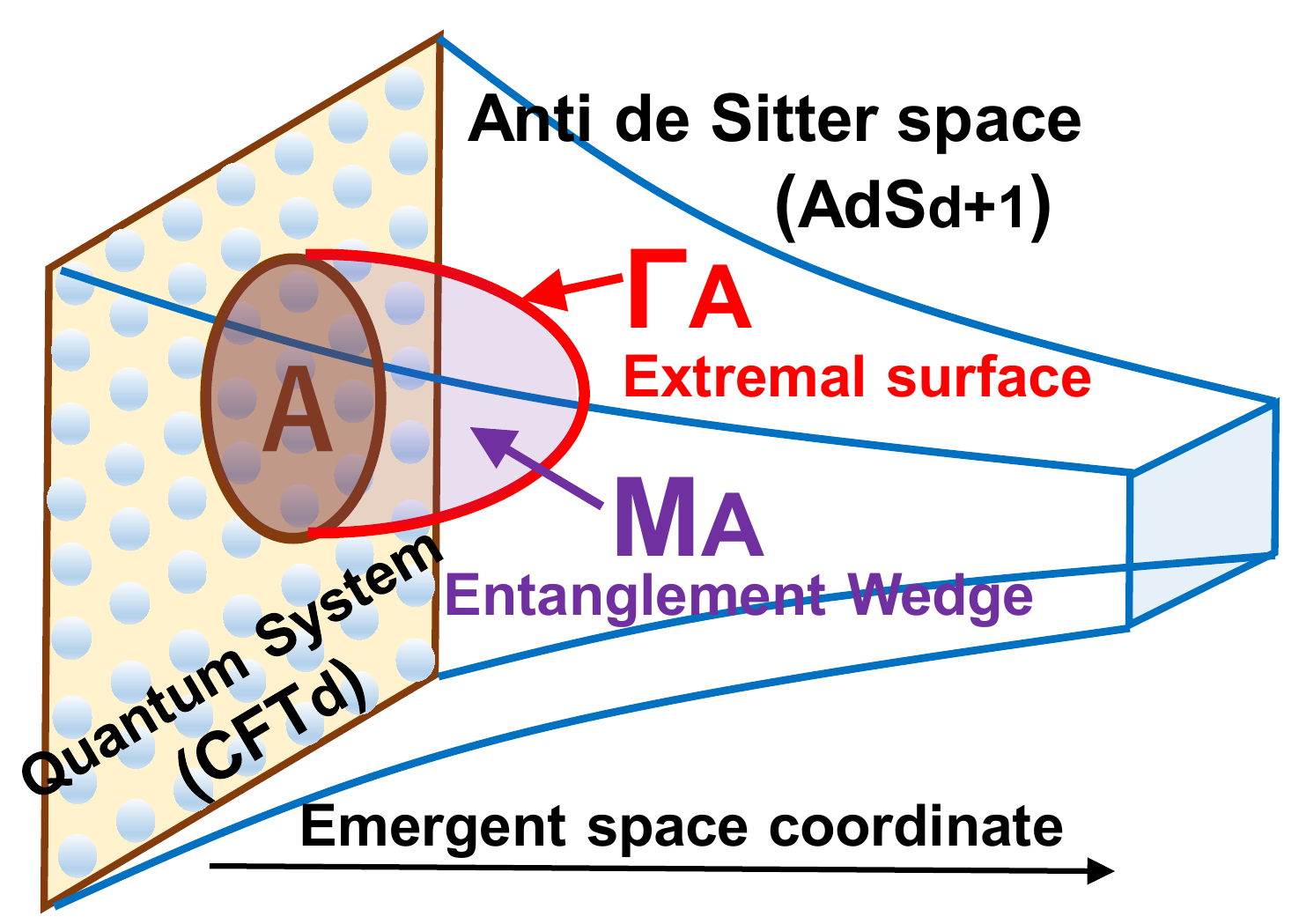}
   \caption{Sketches of AdS/CFT and holographic entanglement entropy.}
   \label{fig:AdS}
\end{figure}

These unusual properties of gravity and black holes get promoted into the idea of {\it gravitational holography} \cite{tHooft:1993dmi,Susskind:1994vu}. Holography argues that a $d+1$ dimensional (quantum) gravity on a spacetime $N_{d+1}$ is equivalent to a nongravitational quantum many-body system in its $d$ dimensional boundary $(\de N)_{d}$, which is often described by quantum field theories. Since the area in the $d+1$ dimensional spacetime looks like a volume in the $d$ dimensional boundary, this is consistent with the area law (\ref{BH}). Moreover, this gives a heuristic understanding of the quantum origin of black hole entropy. More concrete examples have been obtained in string theory. The most well-established example of holography is the {\it Anti-de Sitter/Conformal Field Theory (AdS/CFT) correspondence} \cite{Maldacena:1997re,Gubser:1998bc,Witten:1998qj}, also called {\it gauge/gravity duality} in more general contexts. This argues that the $d+1$ dimensional anti-de Sitter space (AdS$_{d+1}$) is equivalent to a conformal field theory on the $d$ dimensional space (CFT$_{d}$), as depicted in Fig.\ref{fig:AdS}. An intuitive reason for this correspondence is the matching of the symmetry on both sides. The AdS$_{d+1}$ geometry exhibits the isometry $SO(2,d)$, which coincides with the conformal symmetry of CFT$_d$. The AdS/CFT correspondence predicts that physical quantities such as partition functions and correlation functions precisely match between them.

The most famous example at $d=4$ is the equivalence between the $\mathcal N=4$ super Yang-Mills gauge theory in four dimensions and the type IIB string theory on AdS$_5\times$ S$^5$. If we write the rank of the gauge group in the former theory as $N$, the Newton constant $G_N$ in the dual AdS gravity scales as $G_N\sim N^{-2}$. If we turn to $d=2$, the microscopic calculation of entropies of extremal black holes pioneered by \cite{Strominger:1996sh} can also be regarded as an example of AdS$_3/$CFT$_2$. AdS/CFT can also be generalized to asymptotically AdS geometries such that an AdS deformed by gravitational waves is dual to an excited state in a CFT. In this way, AdS/CFT connects the dynamics of quantum gravity, which are challenging to understand directly in general, to those of quantum field theories, which instead can be studied even at nonperturbative levels. For all these reasons, holography has played and will continue to play a key role in advancing our understanding of quantum gravity. 

Our understanding of holography is still far from complete, partly because we do not have any complete proof of the AdS/CFT correspondence and partly because extending holography to more general spacetimes beyond anti-de Sitter spaces is not straightforward. In this Essay, I  will discuss how, to resolve these important issues, it is essential to delve deeper into the fundamental mechanisms of holography by closely re-examining AdS/CFT and its possible generalizations. The tremendous developments on this subject have shown how quantum information theory offers crucial hints to understand how gravity emerges from quantum field theories within the framework of holography. Specifically, various quantities in quantum information theory can capture emergent geometries hidden inside the quantum many-body systems, which eventually become those of gravitational spacetimes. This research direction is gaining momentum and will be further explored alongside the rapid developments in quantum information science.

{\it Holographic entanglement and beyond.}---
One basic question about AdS/CFT is which region in the bulk AdS$_{d+1}$ is dual to a given region $A$ in the CFT$_{d}$. To make this question sharper, we can ask how we can measure the amount of quantum information included in a certain region $A$ on a time slice of the CFT. The best measure for this is the {\it entanglement entropy}, $S_A$, which measures the amount of quantum entanglement between $A$ and its complement $\bar{A}$ for any pure states. Thus, it provides an estimate of the amount of quantum information hidden in $\bar{A}$, defined by $S_A=-\mbox{Tr}[\rho_A\log \rho_A]$, where $\rho_A$ is the reduced density matrix defined by $\rho_A=\mbox{Tr}_{\bar{A}}|\Psi\lb\la\Psi|$ in terms of the total quantum state $|\Psi\lb$. In AdS/CFT, we can compute $S_A$ geometrically by calculating the area of the extremal surface $\Gamma_A$ in an asymptotically AdS geometry which ends on the boundary of $A$ at the AdS boundary (i.e. $\de A=\de \Gamma_A)$ \cite{Ryu:2006bv,Ryu:2006ef,Hubeny:2007xt}. This is the {\it holographic entanglement entropy} given by:
\be
S_{A}=\mbox{Ext}_{\Gamma_A}\left[\frac{\mbox{A}(\Gamma_A)}{4G_N}\right],  \label{ES}
\ee
as explained in Fig.\ref{fig:AdS}. The direct CFT calculations \cite{Holzhey:1994we,Calabrese:2004eu,Nishioka:2018khk} perfectly match with Eq. (\ref{ES}), and the area law of entanglement entropy \cite{Bombelli:1986rw,Srednicki:1993im} follows automatically. This formula, also derived from the bulk to boundary relation in AdS/CFT \cite{Casini:2011kv,Lewkowycz:2013nqa}, manifestly represents the nature of holography, namely, the fact that the algebraically complicated properties in quantum many-body systems are transformed into a simpler geometric one in the dual gravitational spacetimes, highlighted by the holographic derivation of strong subadditivity \cite{Headrick:2007km,Wall:2012uf}.

If we consider a black hole in AdS and take the subsystem $A$ to be the total region, then the holographic entanglement entropy, $S_{A}$, is reduced to the black hole entropy, $S_{BH}$. It would be helpful, therefore, to revisit the concept of black hole entropy from this modern viewpoint. First, we have to stress that in the absence of a complete proof of holography, we still do not know how to compute black hole entropy directly based on quantum gravity. A holographic consideration implies that huge entropy $O(N^2)$ arises from the entanglement between inside and outside of the black hole. However, we know that the degrees of gravity are $O(1)$, i.e., just the gravitons and several other matter fields in general relativity. How can we explain such a large entropy from them? Holographic entanglement suggests that a new degree of freedom emerges when we divide a gravitational space into two parts along a surface $\Gamma_A$. This is much like the appearance of quarks in QCD by cutting a meson into two parts. Such an emergent degree of freedom is called {\it gravity edge modes}. \cite{Donnelly:2016auv,Freidel:2020xyx,Freidel:2023bnj} as it is analogous to those in the topological materials. In other words, gravity is in a sort of confining phase, and the division into two subregions deconfines gravity. This may suggest that a graviton is a bound state of some fundamental degrees of freedom. Indeed, string theory implies that we can view closed strings, whose low-energy excitations are gravitons, as being equivalently described by open strings, whose low-energy modes are gauge fields. This qualitatively estimates entanglement entropy in string theory across a horizon \cite{Susskind:1994sm,Lin:2017uzr}, which can explain why black hole entropy becomes $O(N^2)$. A precise calculation has not been performed yet, but may become possible by developing computations of entanglement entropy in string theory \cite{Dabholkar:2001if,He:2014gva,Witten:2018xfj,Takayanagi:2019tvn,Dabholkar:2022mxo,Dabholkar:2023ows}.

{\it Quantum corrections and entanglement wedges.}---The area formula for holographic entanglement entropy, Eq. (\ref{ES}), describes the leading classical contributions $O(G_N^{-1})=O(N^2)$ in gravity theory. When quantum corrections are incorporated perturbatively, we obtain the {\it quantum extremal surface formula} \cite{Faulkner:2013ana,Engelhardt:2014gca}:
\be
S^{(q)}_{A}=\mbox{Ext}_{\Gamma_A}\left[\frac{\mbox{A}(\Gamma_A)}{4G_N}+S^{(G)}_{M_A}\right].  \label{QES}
\ee
Here $M_A$ is the entanglement wedge and denotes the region in the bulk AdS surrounded by the surface $\Gamma_A$, known as the {\it quantum extremal surface}, and the AdS boundary. In the above formula, we vary $\Gamma_A$ and extremalize the full functional given by the area term plus the bulk entanglement entropy $S^{(G)}_{M_A}$ for the subregion $M_A$ in the effective field theory of the gravity on AdS$_{d+1}$. From this analysis, one finds that the information in the subregion $A$ in CFT$_d$ corresponds to that in the entanglement wedge $M_A$ in bulk gravity, known as {\it entanglement wedge reconstruction} \cite{Czech:2012bh,Wall:2012uf,Headrick:2014cta,Jafferis:2015del}. This provides an answer to the question we asked in the beginning, namely, which region in the bulk AdS$_{d+1}$ corresponds to a given subregion $A$. However, Eq. \ref{QES} is highly reliant on the effective field theory description, which limits our ability to compute the nonperturbative corrections. Therefore, it would be desirable to compute the full holographic entanglement entropy by examining the solvable CFT dual in, for example, the zero gauge coupling limit, and reinterpret it in terms of quantum gravity contributions. This approach may reveal a nonperturbative aspect of holographic entanglement entropy. 

A remarkable generalization of Eq.(\ref{QES}) is the {\it island formula} \cite{Penington:2019npb,Almheiri:2019psf}: 
\be
S^{(I)}_{A}=\mbox{Ext}_{\Sigma}\left[\frac{\mbox{A}(\Sigma)}{4G_N}+S^{(G)}_{A\cup \Sigma}\right],  \label{Is}
\ee
that computes the entanglement entropy of a subregion $A$ in a quantum field theory on a spacetime when it is interacting along an interface with another spacetime where gravity exists. This new region, $\Sigma$, is called {\it island} and is initially chosen as a subregion inside the gravitating spacetime which extremalizes the above functional. This shows how gravitational interactions can significantly alter the calculation of entanglement entropy, even from a certain distance away from $A$.

Interesting considerations arise when applying the island formula to an evaporating black hole. It is known that a black hole emits Hawking radiation, losing its mass, and finally evaporates completely. In this process, we encounter the fundamental problem that the huge amount of information originally hidden in the black hole may be lost after its evaporation, contradicting the unitarity of quantum mechanics, for which information should be conserved. This is the well-known {\it black hole information problem}. The island formula predicts that the entanglement entropy between the black hole and radiation follows a special behavior called the Page curve \cite{Page:1993wv}, where the entanglement entropy vanishes as the black hole evaporates. This argument would imply that the information is not lost under evaporation \cite{Penington:2019npb,Almheiri:2019psf}, rather transferred from the black hole to its emitted radiation, resolving the paradox.

A useful tool for exploring the paradox and deriving the island formula for the entanglement entropy is the replica wormhole, a complex spacetime configuration, connecting the subregion $A$ with the island \cite{Penington:2019kki,Almheiri:2019qdq}. For two-dimensional Euclidean gravity theories, this wormhole is one of the saddle solutions of the quantum-corrected Einstein equation and becomes the dominant solution when the island emerges. In the calculation of entanglement entropy, this wormhole cancels the entanglement between $A$ and the island. This significantly reduces the entanglement entropy and eventually leads to the Page curve. However, in Lorentzian spacetime, where we have the real-time evolution, we still do not fully understand why the island emerges in the calculation of entanglement entropy. In other words, how can we observe such a wormhole in dynamical spacetime that includes an evaporating black hole?

This important question is directly related to what the Hilbert space in quantum gravity looks like. Though the conventional idea in quantum field theories argues that $A$ and $\Sigma$, which are space-like separated, are independent, the island formula says that they are in the same Hilbert space due to the presence of gravity. As explained in \cite{Almheiri:2019qdq, Suzuki:2022xwv}, one clue supporting this idea is the connection between the island formula Eq.(\ref{Is}) and the prescription given in \cite{Takayanagi:2011zk,Fujita:2011fp} for computing the holographic entanglement entropy in {\it AdS/BCFT}, i.e., the holographic dual of a CFT on a space with boundaries. In AdS/BCFT, the gravity dual of a $d$-dimensional BCFT is given by a part of AdS$_{d+1}$ which is surrounded by a surface in the bulk AdS, called the {\it end-of-the-world brane} (EOW brane). In the context of brane-world theory \cite{Randall:1999ee,Randall:1999vf}, a world brane is a higher-dimensional surface in a larger spacetime where our universe, with its four dimensions, is confined. An EOW brane represents a boundary or a cutoff where the spacetime effectively terminates or changes significantly. In this case, when computing the holographic entanglement entropy, we allow the extremal surface $\Gamma_A$ to end at the EOW brane \cite{Takayanagi:2011zk,Fujita:2011fp}, and thus we extremalize the area of $\Gamma_A$ by changing the ending geometry on the brane, which is identified with the island $\Sigma$. To make this connection precise, we need to better understand the generalization of AdS/CFT to the brane-world \cite{Randall:1999ee,Randall:1999vf,Gubser:1999vj,Karch:2000ct}. Exploring {\it brane world holography} will bring us important insights into the degrees of freedom in quantum gravity. 

{\it More general measures of quantum information.}---Entanglement entropy quantifies quantum entanglement in a bipartite quantum system, but only for a pure state. We could gain a deeper understanding of holography if we could find holographic formulas to compute the amount of entanglement in mixed states, as well as multipartite entanglement. For the entanglement of mixed states, it was suggested that the squashed entanglement \cite{SqE}, which is expected to be the most ideal measure of entanglement, is equivalent to half of the mutual information \cite{Hayden:2011ag}. However, it is important to note that there are infinitely many measures of entanglement for mixed states \cite{Horodecki:2009zz}. Therefore, it is crucial to develop a comprehensive understanding of holographic computations for these various measures. For this purpose, it would be helpful to implement within the context of holography operational procedures, such as local operations and classical communication \cite{Horodecki:2009zz}. This would allow us to identify the geometric dual of each measure, since entanglement measures in quantum information theory are defined operationally.

For a correlation measure, the minimal cross-section of the entanglement wedge was argued to be equal to the {\it entanglement of purification} \cite{Takayanagi:2017knl,Nguyen:2017yqw}. For a given mixed state, we can extend the Hilbert space and describe it as a pure state, the so-called purification. The entanglement of purification is defined as the minimum of entanglement entropy among all possible purifications \cite{EoP}. Since these measures in quantum information theory typically involve a minimization over huge auxiliary Hilbert spaces, practical calculations in quantum field theories turn out to be very difficult. Nevertheless, by choosing special CFTs and taking advantage of the conformal symmetries and maximal chaotic property, we may be able to overcome this computational difficulty.  Though we still do not know the best measure for multipartite entanglement, recently an interesting candidate called {\it multi-entropy} was proposed \cite{Gadde:2022cqi,Penington:2022dhr,Gadde:2023zzj}, defined by considering some copies of multipartite density matrices and by contracting their indices following a certain rule of permutations. This is computable via replica methods in CFTs \cite{Harper:2024ker,Gadde:2024taa} and can potentially have geometrical formulas via holography based on junctions of extremal surfaces. Taking this as an important hint, we may find genuine multipartite entanglement measures with manifest gravity duals, which would tell us how multipartite entanglement is related to spacetime geometry.

{\it Emergent spacetime from entanglement.}---The area formula, Eq. (\ref{ES}), suggests that for every Planck-scale area in gravitational spacetime, a Bell pair of qubits exists on the surface $\Gamma_A$. Since the choice of $A$ is arbitrary, by covering the whole spacetime by extremal surfaces, we may expect that the spacetime is filled with infinitely many Bell pairs. We can interpret gravitational waves in the AdS as dynamical evolutions of quantum entanglement in qubits. For example, the perturbative Einstein equation in the latter can be understood as a basic dynamics of entanglement entropy, akin to the first law of thermodynamics \cite{Blanco:2013joa,Lashkari:2013koa,Faulkner:2013ica}. This implies that the microscopic structure of spacetime in gravity may be regarded as the collection of entangled qubits \cite{Swingle:2009bg,VanRaamsdonk:2010pw}. These considerations raise a very fundamental question: How can we concretely describe the emergence of gravity from quantum information so that we can formulate quantum gravity based on this principle?  

A useful insight for addressing this problem is the observation that tensor networks can serve as toy models of AdS/CFT. For example, the {\it Multiscale Entanglement Renormalization Ansatz} (MERA) \cite{Vidal:2007hda} describes the ground states at a quantum critical point \cite{Swingle:2009bg}. A notable feature of tensor network description is that the entanglement entropy, $S_A$, in a quantum many-body system can be estimated from the network geometry by counting the minimal number of linking bonds surrounding the subregion $A$. This estimation aligns qualitatively with holographic calculations (\ref{ES}). Currently, MERA seems to fit nicely with the light-like slices of AdS geometries, as these slices consist of unitary or isometry gates and describe the Lorentzian time evolution \cite{Beny:2011vh,Czech:2015kbp}. On the other hand, a time slice of AdS corresponds to a network representing a Euclidean path-integral, which can be formulated by using the path-integral optimization \cite{Caputa:2017urj,Caputa:2017yrh}. Additionally, there have been other interesting approaches to tensor network models of holography that consider quantum error correcting codes \cite{Pastawski:2015qua} and random tensors \cite{Hayden:2016cfa}, where the minimal area formula, Eq. (\ref{ES}), becomes exact. 

However, we still lack a precise understanding of how a genuine gravitational spacetime is described by a tensor network or its generalization. One crucial unsolved problem is that holographic tensor network models have not yet leveraged the special feature of holographic CFTs, namely, strongly coupled CFTs. This strong coupling dynamics is crucial for explaining locality in bulk AdS gravity at scales much smaller than the AdS radius \cite{Heemskerk:2009pn}. To address this, one strategy may be to examine explicit examples of AdS/CFT in string theory, such as \cite{Eberhardt:2018ouy}, where both the CFT and string theory sides are solvable. 

Another direction is to take into account the dynamics of gauge theory in the tensor networks. For this purpose, string theory provides a heuristic picture. In string theory, the most microscopic constituents of matter are considered to be closed and open strings. At low energy, a closed and open string describe a graviton and a Yang-Mills gauge field, respectively. Since an open and closed string are two different ways to view the same string world-sheet, we expect they describe two equivalent theories, which is called open-closed duality and is expected to be a fundamental reason why we expect the AdS/CFT. Therefore, we may be able to find the emergence of gravity (i.e., closed strings) from the color degrees of freedom (i.e., open strings) in the tensor network description of holography. 

A related question is understanding how internal spaces, such as S$^5$, emerge in the AdS$_5\times$S$^5$ geometry, from a quantum information perspective. Currently, it is unclear whether an area in these internal spaces can be interpreted as a specific quantity in quantum information theory, though it was suggested that it is related to the entanglement entropy in field spaces \cite{Mollabashi:2014qfa}. A more ambitious and fundamental question is: How does the diffeomorphism invariance of gravity emerge from quantum information? We anticipate that examining explicit examples from string theory will aid in addressing these questions in the coming years. Additionally, interpreting AdS/CFT in terms of renormalization group flows \cite{deBoer:1999tgo,Lee:2013dln} and TTbar deformations \cite{McGough:2016lol} will also provide important clues.

{\it Gravity as a quantum computer.}---Thinking of tensor networks as a specific type of quantum circuits, we may be able to dig deeper into the fundamental mechanisms of holography by considering AdS/CFT as a quantum computer \cite{Swingle:2009bg,Milsted:2018san,Takayanagi:2018pml}. This perspective allows us to leverage the extensive advancements made in the field of quantum computing. Another motivation for this approach is the expectation that gravitational dynamics correspond to those of the fastest quantum computers via holographic duality \cite{Lloyd_2000,Maldacena:2015waa}, because the CFT dual to the classical AdS gravity is known to be strongly coupled. Therefore, it will be vital to determine what kind of quantum computer corresponds to AdS/CFT in the coming years. This could provide useful hints for developing new quantum computers. On the other hand, if we can learn the basic mechanisms of holography through this analysis, we may be able to extend the principles of holography to more general spacetimes beyond AdS. For this purpose, it might be useful to refer to earlier proposals in \cite{Miyaji:2015yva,Bousso:2022hlz,Bousso:2023sya,Balasubramanian:2023dpj,Gupta:2025jlq} on holography for general spacetimes from the viewpoint of quantum entanglement.

In quantum computing, the difficulty of completing a computational task is measured by {\it computational complexity} \cite{Aaronson:2016vto}. In AdS/CFT, the amount of computational complexity for generating a given quantum state was conjectured to be estimated by the volume of the maximal time slice \cite{Roberts:2014isa}. There is also a similar but different proposal that is given by the gravity on-shell action in the Wheeler-DeWitt patch \cite{Roberts:2014isa,Brown:2015bva}, defined by the union of all spatial slices anchored at a given boundary time.
However, it was pointed out in \cite{Bouland:2019pvu,Aaronson:2022flk} that AdS/CFT appears to perform tasks that exceed what quantum computers are expected to manage. On the other hand, in the field of quantum computing, it is widely believed that quantum computers can efficiently simulate all physical processes, including quantum gravity; this is known as the {\it quantum extended Church-Turing thesis}. Thus, one might worry that the performance of AdS/CFT may contradict this reasonable belief \cite{Bouland:2019pvu,Aaronson:2022flk}. For instance, measuring the complexity of a generic state in quantum many-body systems is considered quite challenging, as this cannot be accomplished within a polynomial time using quantum computers. A measurement of entanglement entropy is expected to be similarly difficult, so that quantum computers are unable to perform it within polynomial time \cite{Gheorghiu:2020sko,Aaronson:2022flk}. Even the estimation of the ground state energy of a generic quantum many-body system, the so-called {\it local Hamiltonian problem}, is known to be QMA-complete \cite{Kitaev+02}, hence more difficult than what efficient quantum computers can handle.

We also need to be cautious that the calculations in AdS/CFT get simplified only in its classical gravity approximation, which corresponds to the large $N$ limit of CFT. This implies the possibility that estimating physical quantities that behave like $O(N^2)$ in the large $N$ limit may become drastically easier if we are tolerant about $O(1)$ errors. Moreover, in AdS/CFT and quantum field theories, ground states are usually generated via Euclidean path integrals rather than Lorentzian ones. Thus, the calculations done in AdS/CFT, especially with classical solutions, may involve both unitary and non-unitary gates, as in the path-integral optimization procedures \cite{Caputa:2017urj,Caputa:2017yrh}. Indeed, as shown in \cite{cade}, if we assume the existence of a guided state with a large overlap with the correct ground state (know as the {\it guided local Hamiltonian problem} \cite{Gharibian_2023}) and we allow $O(1)$ errors in the estimation, then calculating the ground state energy falls into the complexity class BPP, namely a task that can be performed efficiently even by classical computers. The implications for holography may be heuristic, but will motivate us to examine what  AdS/CFT does in the language of quantum circuits when calculating various quantities such as the correlation functions and entanglement entropy in CFTs through gravity analysis.

{\it Pseudo-entropy and emergent time.}---Holographic entanglement entropy suggests that the extra space-like coordinate in AdS may emerge from quantum entanglement in its dual CFT, which is highlighted by the tensor network framework. This leads to a significant question: Can the time coordinate in holographic spacetimes also emerge from some degrees of freedom in quantum information?  One potential clue to this fundamental question could be the extension of holographic entanglement entropy, Eq. (\ref{ES}), to scenarios where the initial state $|\Psi_i\lb$ and final state $|\Psi_f\lb$ are different. While entanglement entropy is typically defined for a single quantum state by focusing on a subsystem $A$, we can generalize this concept to the case where the initial state $|\Psi_i\lb$ and final state $|\Psi_f\lb$ are different. Introducing the reduced transition matrix $\tau_A$:
\ba
\tau_A=\mbox{Tr}_{\bar{A}}\frac{|\Psi_i\lb\la\Psi_f|}{\la \Psi_f|\Psi_i\lb},
\ea
we can define a von-Neumann-entropy-like quantity, called {\it pseudo-entropy} \cite{Nakata:2020luh}:
\ba
S^{(p)}_A=-\mbox{Tr}[\tau_A\log\tau_A].
\ea
Since $\tau_A$ is not Hermitian, the pseudo-entropy can take complex values in general. The quantum information interpretation of the imaginary part of the pseudo-entropy is an important open problem, related to the emergence of time in cosmological holography, as we will discuss later.

We can realize a scenario where the initial and final states differ by considering a Euclidean (or imaginary) time-dependent CFT. This setup is dual to a Euclidean time-dependent AdS geometry via AdS/CFT. Interestingly, in this case, we can again compute geometrically the pseudo-entropy using the same formula, Eq. (\ref{ES}), so holographic pseudo-entropy is given by the area of the minimal surface divided by $4G_N$ \cite{Nakata:2020luh}. In this Euclidean case, for real-valued metrics, the holographic pseudo-entropy becomes real and non-negative. Moreover, we can also compute geometrically the pseudo-entropy in Lorentzian AdS/CFT setups, where we often encounter its imaginary part. One example is given by the {\it time-like entanglement entropy} \cite{Doi:2022iyj,Doi:2023zaf} where we choose the subsystem $A$ to extend in the time direction. This quantity can be computed via an analytical continuation of the standard entanglement entropy where $A$ is space-like.  In this case, since the subsystem $A$ is not on a regular time slice, there is a causal influence among points on $A$, and we cannot decompose the Hilbert space into $A$ and $\bar{A}$, unlike in the case of entanglement entropy.
This causal influence makes the density matrix non-Hermitian \cite{Parzygnat:2022pax,Kawamoto:2025oko,Milekhin:2025ycm}, so the pseudo-entropy can assume complex values. In AdS/CFT, the imaginary part arises because the extremal surface $\Gamma_A$ can become partly time-like \cite{Doi:2022iyj,Doi:2023zaf,Heller:2024whi}. This implies that the imaginary part of pseudo-entropy may be linked to the emergence of the time coordinate in holography. Therefore, further investigations into time-like entanglement, such as from the perspective of quantum circuits and string theory analysis, will provide deeper insights into how holography works. Refer also to \cite{Carignano:2024jxb,Bou-Comas:2024pxf} for measurement of this quantity in quantum simulators.

Another promising setup to study involves {\it AdS traversable wormholes}. An AdS wormhole connects two asymptotically AdS regions through a throat region. When we can send a signal from one AdS boundary to the other, we say that it is traversable. We can show that the traversability of AdS wormhole is associated with the non-Hermitian properties of the density matrix, which leads to the imaginary part of pseudo-entropy \cite{Kawamoto:2025oko}.
\begin{figure}[ttt]
   \centering
   \includegraphics[width=6cm]{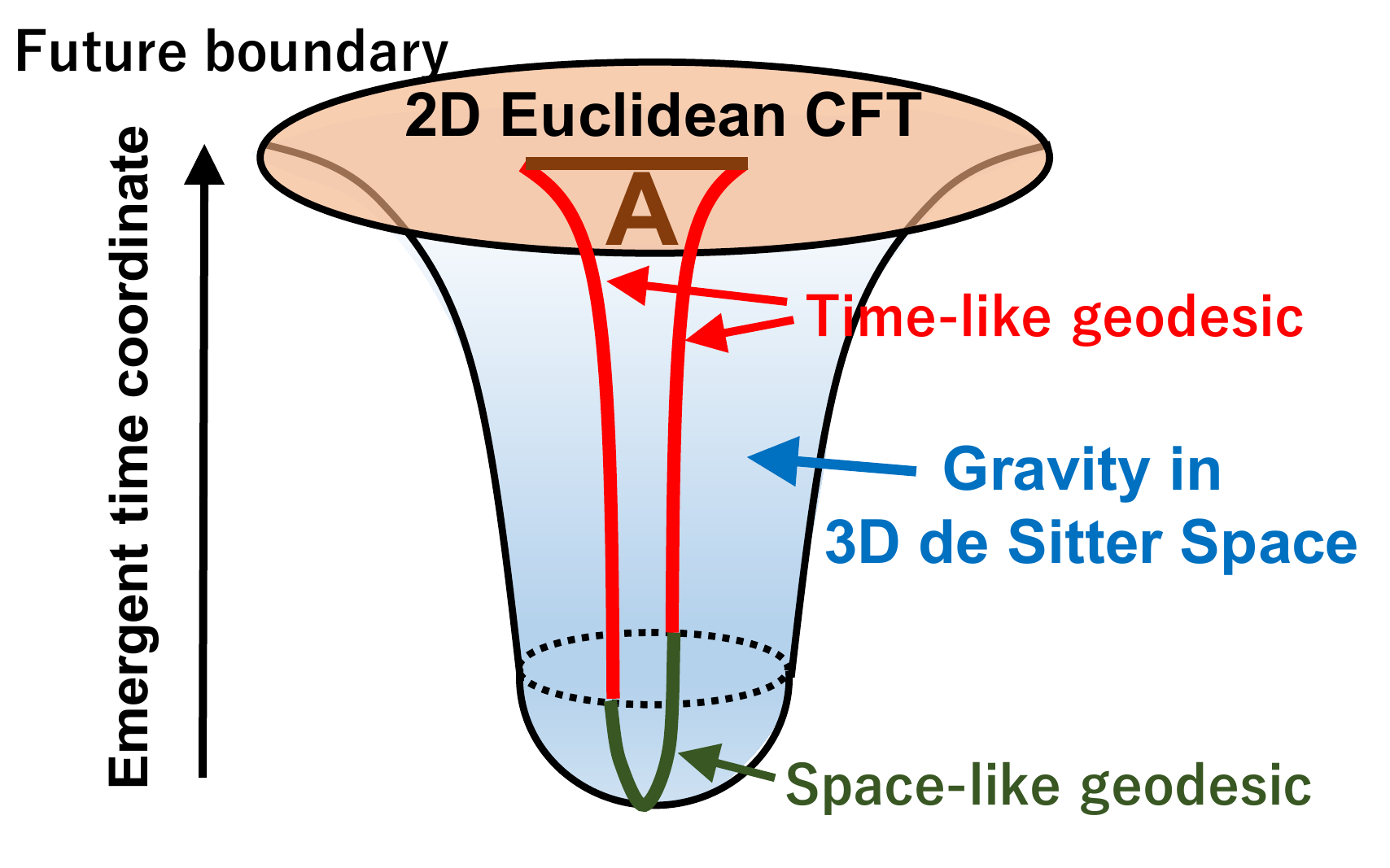}
   \caption{Sketches of dS$_3/$CFT$_2$ and holographic pseudo entropy. The union of red and green geodesics is $\Gamma_A$.}
   \label{fig:dS}
\end{figure}

The mechanism of emergent time is particularly crucial when considering holography in cosmological spacetimes, such as de Sitter spaces. In these cases, the boundary of a de Sitter space is situated at both future and past infinity. The holography framework in de Sitter spaces, known as {\it dS/CFT} \cite{Strominger:1996sh,Maldacena:2002vr}, argues that gravity on a $d+1$ dimensional de Sitter space (dS$_{d+1}$) is dual to a $d$ dimensional Euclidean CFT (CFT$_d$), which exists at future infinity. This assumption relies on the Hartle-Hawking state so that a Euclidean instanton creates the initial state, as depicted in Fig.\ref{fig:dS}. However, understanding dS/CFT is considerably more challenging than its counterpart, AdS/CFT, partly because the dual CFTs are expected to be non-unitary, as shown in explicit examples \cite{Anninos:2011ui,Maldacena:2019cbz,Cotler:2019dcj,Cotler:2019nbi,Hikida:2021ese}, and partly because the time coordinate should somehow emerge from the Euclidean CFT. Exploring holography in these cosmological spacetimes is crucial for advancing our understanding of the origin of the Universe. As recent developments have provided many relevant insights and implications, this problem might be addressed in the coming years by combining CFT computations with quantum information considerations.

For example, in dS$_3/$CFT$_2$, we expect that the central charge of the dual CFT becomes imaginary \cite{Maldacena:2002vr,Hikida:2021ese} in the classical gravity limit. This scenario can be described by a Liouville CFT \cite{Hikida:2022ltr,Narovlansky:2023lfz,Verlinde:2024zrh,Collier:2024kmo}. It is natural to believe that this non-unitary nature is the reason for the emergence of time, which is missing in  AdS/CFT scenarios. Consequently, the density matrix in the dual CFT becomes non-Hermitian, suggesting that entanglement entropy should be regarded as pseudo-entropy. The holographic pseudo-entropy in dS$_3/$CFT$_2$ can be derived from the geodesic length which connects two points in the future infinity, as illustrated in Fig.\ref{fig:dS}. This length turns out to be complex-valued as a part of this geodesic is time-like \cite{Doi:2022iyj,Narayan:2022afv,Doi:2023zaf}, whose imaginary part is proportional to the imaginary central charge. This suggests that the time coordinate in dS$_3$ may emerge from the imaginary part of pseudo-entropy, while its real part explains the emergent space coordinate, as for holographic entanglement entropy. In this way,  pseudo-entropy seems to capture the spacetime structure of the holographic Universe. The classical geometry underlying general relativity breaks down when quantum gravity effects become dominant, such as during the Universe's creation in the Big Bang. Nevertheless, entanglement entropy and pseudo-entropy are expected to remain valid in the quantum many-body system dual to such a quantum Universe. In the upcoming years, we may be able to use these quantities to formulate the degrees of freedom in quantum gravity.

{\it Concluding remarks.}---Before concluding this Essay, we would like to emphasize the importance of developing and testing theoretical ideas regarding the emergence of gravity from quantum information. Thinking of recent tremendous developments of various quantum simulators, this can be achieved by realizing {\it mini-Universes} in condensed matter experiments. Entanglement entropy has been measured already, for example, in cold atom systems \cite{Islam_2015}, trapped ions \cite{Brydges_2019,Joshi_2023}, and superconducting quantum processors \cite{Satzinger_2021}. It would be fascinating to construct a toy model of holography in real experiments using, for example, quantum Hall effects, where we can probe both the boundary and bulk physics \cite{Hotta:2022aiv,PhysRevResearch.4.L012040,france2025electricallyinducedbulkedge}.

Surprising connections between holographic duality and quantum information theory have recently provided tremendous developments to uncover various important aspects of quantum gravity. However, this seems to be just the tip of the iceberg. In the near future, by exploring these connections from various new angles, including those suggested in this Essay, we hope to achieve a more comprehensive understanding of quantum gravity so that we can access fundamental and realistic problems, such as the creation of the Universe.

{\it Acknowledgements:} 

TT is very grateful to Ryu Hayakawa and Tomoyuki Morimae for valuable discussions on the connection between complexity and holography and for helpful comments on this draft. TT also thanks Pawel Caputa, Jonathan Harper, Alexander Jahn, Shinsei Ryu for useful conversations when this draft was prepared. This work is supported by MEXT KAKENHI Grant-in-Aid for Transformative Research Areas (A) through the ``Extreme Universe'' collaboration: Grant Number 21H05182, 21H05187, and by JSPS Grant-in-Aid for Scientific Research (B) No.~25K01000.


\bibliography{essay.bib}


\end{document}